\documentclass[12pt,psfig]{article}
\makeatletter
\def\@cite#1#2{{[{#1}]\if@tempswa\typeout
{IJCGA warning: optional citation argument
ignored: `#2'} \fi}}

\newcount\@tempcntc
\def\@citex[#1]#2{\if@filesw\immediate\write\@auxout{\string\citation{#2}}\fi
  \@tempcnta\z@\@tempcntb\m@ne\def\@citea{}\@cite{\@for\@citeb:=#2\do
    {\@ifundefined
       {b@\@citeb}{\@citeo\@tempcntb\m@ne\@citea\def\@citea{,}{\bf ?}\@warning
       {Citation `\@citeb' on page \thepage \space undefined}}%
    {\setbox\z@\hbox{\global\@tempcntc0\csname b@\@citeb\endcsname\relax}%
     \ifnum\@tempcntc=\z@ \@citeo\@tempcntb\m@ne
       \@citea\def\@citea{,}\hbox{\csname b@\@citeb\endcsname}%
     \else
      \advance\@tempcntb\@ne
      \ifnum\@tempcntb=\@tempcntc
      \else\advance\@tempcntb\m@ne\@citeo
      \@tempcnta\@tempcntc\@tempcntb\@tempcntc\fi\fi}}\@citeo}{#1}}
\def\@citeo{\ifnum\@tempcnta>\@tempcntb\else\@citea\def\@citea{,}%
  \ifnum\@tempcnta=\@tempcntb\the\@tempcnta\else
   {\advance\@tempcnta\@ne\ifnum\@tempcnta=\@tempcntb \else
\def\@citea{--}\fi
    \advance\@tempcnta\m@ne\the\@tempcnta\@citea\the\@tempcntb}\fi\fi}
\makeatother

\newcommand{\gsim}{\lower.7ex\hbox{$\;\stackrel{\textstyle>}{\sim}\;$}}
\newcommand{\lsim}{\lower.7ex\hbox{$\;\stackrel{\textstyle<}{\sim}\;$}}
\newcommand{\be}{\begin{equation}}
\newcommand{\ee}{\end{equation}}
\newcommand{\bea}{\begin{eqnarray}}
\newcommand{\eea}{\end{eqnarray}}

\usepackage{amssymb}

\def\baselinestretch{1}
\begin{document}
\catcode`@=11
\newtoks\@stequation
\def\subequations{\refstepcounter{equation}%
\edef\@savedequation{\the\c@equation}%
  \@stequation=\expandafter{\theequation}
  \edef\@savedtheequation{\the\@stequation}
  \edef\oldtheequation{\theequation}%
  \setcounter{equation}{0}%
  \def\theequation{\oldtheequation\alph{equation}}}
\def\endsubequations{\setcounter{equation}{\@savedequation}%
  \@stequation=\expandafter{\@savedtheequation}%
  \edef\theequation{\the\@stequation}\global\@ignoretrue

\noindent}
\catcode`@=12
\begin{titlepage}
\title{{\bf
Higher codimension braneworlds from intersecting branes}}
\vskip2in
\author{
{\bf Ignacio Navarro$$\footnote{\baselineskip=16pt E-mail: {\tt
ignacio.navarro@durham.ac.uk}}}
$\;\;$and$\;\;$
{\bf Jose Santiago$$\footnote{\baselineskip=16pt E-mail: {\tt
jose.santiago-perez@durham.ac.uk}}}
\hspace{3cm}\\
 $$~{\small IPPP, University of Durham, DH1 3LE Durham, UK}.
}

\date{}
\maketitle
\def\baselinestretch{1.15}
\begin{abstract}
\noindent 

We study the matching conditions of intersecting brane worlds in
Lovelock gravity in arbitrary dimension. We show that intersecting
various codimension 1 and/or codimension 2 branes one can find
solutions that represent energy-momentum densities localized in the
intersection, providing thus the first examples of infinitesimally
thin higher codimension braneworlds that are free of singularities and
where the backreaction of the brane in the background is fully taken
into account. 

\end{abstract}

\thispagestyle{empty} \vspace{5cm}  \leftline{}

\vskip-20cm \rightline{} \rightline{IPPP/04/06}
\rightline{DCPT/04/12} \rightline{hep-th/0402204} \vskip3in

\end{titlepage}
\setcounter{footnote}{0} \setcounter{page}{1}
\newpage
\baselineskip=20pt

\section{Introduction}

Extended objects of arbitrary dimension, the so-called branes, are
nowadays a common ingredient of beyond the Standard Model theories
that hypothesise the existence of extra dimensions in our
universe. These objects can be described in terms of topological
defects of field theories in higher dimensions or can have a more
fundamental character like the D-branes of string
theory. Phenomenological models containing branes have proven to be
useful in attacking many of the main problems of high energy physics
theories. The important property that is exploited in order to
construct these models is that some of the fields can be confined to
the submanifold of spacetime regarded as a brane: the zero modes of
the topological defects or the gauge theories living in the
worldvolume of the string theory D-branes. At low energies, from the
point of view of gravity the brane will yield then a distributional
term in the energy-momentum tensor that has a delta-like behaviour,
$i.e.$ an energy-momentum density localized in some submanifold of the
whole spacetime. It is then of great importance for these models to
find solutions of higher dimensional gravity that correspond to these
configurations. The nature of the solutions found depend crucially on
the codimension of the brane, $i.e.$ the number of extra
dimensions. In fact, in Einstein gravity, one can establish an analogy
between the behaviour of a codimension $n$ brane and solutions
corresponding to a point particle in $n+1$ dimensions. Gravity is
trivial in 1+1 dimensions (the action is a topological invariant)
while in 2+1 a point particle simply produces a conical deficit at its
position \cite{Deser:tn}, the spacetime being flat where there is no
matter in both cases. Analogously, the codimension 1 brane simply
produces a jump in the first derivatives of some metric components at
its position and can be dealt with using the Israel junction
conditions \cite{Israel:rt,Shiromizu:1999wj}, while a codimension 2
brane produces a conical singularity in the transverse space
\cite{Chen:2000at}\footnote{This property makes 6D models where
  observable fields are confined to a 3-brane attractive from the
  point of view of the Cosmological Constant Problem, since selftuning
  ideas can find in them a natural implementation
  \cite{Chen:2000at,Carroll:2003db}.}.  

Actually, the situation is more subtle for codimension 2 branes. One
can find well behaved solutions for a pure tension brane in Einstein
gravity (when the brane energy-momentum tensor is proportional to the
induced metric) just including a deficit angle in the spacetime, but
for a general brane it seemed impossible to find solutions if one
requires a non-singular induced metric on the brane
\cite{Cline:2003ak}. This situation was anticipated in
\cite{Geroch:qn}, where general solutions with distributional sources
were studied in 4D Einstein gravity and it was found that only matter
shells (codimension 1 sources) gave a well behaved solution. A way out
of this problem was proposed in \cite{Bostock:2003cv}, where it was
shown that if one includes a Gauss-Bonnet term in the
action\footnote{In every odd number of dimensions, 2N+1, one can add
  to the action a term of $N$th order in the curvature tensor (the $N$th
  order Euler density) and the equations of motion remain second order
  differential equations for the metric. These Lagrangians are known
  as the Lanczos-Lovelock Lagrangians \cite{Lovelock:1971yv} and are
  believed to arise as the low energy limit of string theory, since
  they are the only ghost-free effective actions for spin two fields
  \cite{Zwiebach:1985uq}.} the matching conditions can be satisfied
for a general brane energy-momentum tensor and, moreover, requiring
regularity of the solution one recovers the lower dimensional Einstein
equations for the induced metric and matter on the brane from the
matching conditions, independently of the bulk solution. 

For $n>2$ the situation is quite different, and one finds
singularities (black holes) in the metric describing point particles
in $n+1$ dimensions (see \cite{Geroch:qn} for a discussion in 4
dimensions). 
In the same fashion, known solutions of higher
codimension branes present singularities at its positions, naked
\cite{Charmousis:2001hg} or surrounded by an event horizon (the black
branes of \cite{Horowitz:cd}). This is the main reason why higher
codimension braneworlds have not been used in the literature to
construct phenomenological models as much as their lower codimension
cousins, since one does not have a well defined submanifold at the
position of the brane. In fact, the usual procedure when dealing with
higher codimension branes is to neglect the effect of the brane on the
background. However, it is then hard to make any prediction about the
nature of the gravity induced on the brane ($i.e.$ how gravity for
brane observers is), since what one has to compute is the effect of
matter on the brane on its own induced metric. In any case, if the
curvature grows as one approaches the brane, the solution is singular
at its position and the singularity is cut-off by a finite brane
width, physical predictions would depend on the brane width and
internal structure. It is therefore interesting to look for solutions that
are not singular at the position of the brane since in this case one
can make unambiguous predictions valid for any microscopic brane
theory. For the codimension 2 case, this can be done when one
considers the Gauss-Bonnet term in the action \cite{Bostock:2003cv},
and it was found that the nature of gravity on the brane does not
yield the result expected from the arguments exhibited in
\cite{Arkani-Hamed:1998rs}, where the self-gravity of the brane is
neglected. 

On the other hand, it has been shown that if one considers the intersection of
two codimension 1 branes in 6D Einstein-Gauss-Bonnet gravity one needs
the presence of a codimension 2 brane with non-zero tension at the
intersection in order to satisfy the maching conditions, due to the
contribution of the Gauss-Bonnet terms \cite{Lee:2004hh}. This is an
interesting observation because the origin of the $\delta^{(2)}$
contribution in the Einstein-Gauss-Bonnet tensor is completely
different from a defict angle, and provides thus another way to find
solutions that represent non-singular codimension 2 branes. In this
paper we generalize these ideas considering the intersection of
various codimension 1 and/or 2 branes. We show that when one considers
the most general theory of gravity in higher dimensions\footnote{The
  only requirement is that the action is torsion free and the
  equations of motion are second order differential equations for the
  metric \cite{Zwiebach:1985uq}.} (the Lagrangian will have the
Einstein-Hilbert term plus the dimensionally continued Euler
densities), one can find non-singular solutions representing branes of
higher codimension living in the intersection of higher dimensional
branes (of lower codimension). 
Branes of codimension up to $N-2$ ($N-1$) in $N$ even (odd) dimensions
can be reproduced in this way with
a non-singular brane induced metric. Intersecting branes are
interesting from the point of view of string phenomenology since one
can build up models with a Standard Model like spectrum
(see \cite{Lust:2004ks} for a recent review). 
Our results indicate that they are also interesting in the sense that one
can build non-singular solutions corresponding to these configurations
where the backreaction of the brane in the background is fully taken
into account even when higher codimension branes are present. 

\section{Brane intersections and matching conditions in Lovelock gravity}

The Lovelock Lagrangian in $D$ dimensions is built up with all the
Euler densities of lower dimensions
\begin{equation}
\mathcal{L}_D=\sum_{p=0}^{\left[\frac{D-1}{2}\right]} \alpha_p
\mathcal{L}_{(p)},
\end{equation}
where $\alpha_p$ is a coefficient of mass dimension $D-2p$ 
and the
square brackets represent the integer part since higher Lovelock terms
are trivial in $D$ dimensions. In particular $\alpha_0$ and $\alpha_1$
represent just a cosmological constant and the higher dimensional
Planck mass, respectively (we will set $\alpha_1=1$ in the following without loss of generality). The order $p$ term is 
\begin{equation}
\mathcal{L}_{(p)}=\frac{1}{2^p} 
\delta^{i_1 \ldots i_{2p}}_{j_1\ldots j_{2p}} 
R^{j_1j_2}_{\phantom{j_1j_2}i_1i_2}
\ldots
R^{j_{2p-1}j_{2p}}_{\phantom{j_{2p-1}j_{2p}}i_{2p-1}i_{2p}},
\end{equation}
where $\delta^{i_1 \ldots i_{2p}}_{j_1\ldots j_{2p}}$
is the Kronecker symbol of order $2p$ and $R^A_{\, BCD}$ is
the $D-$dimensional Riemann tensor.

The resulting equations of motion are
\begin{equation}
\sum_{p=0}^{\left[\frac{D-1}{2}\right]} \alpha_p
G_{(p)AB}=T_{AB},\label{Lovelock:EoM}
\end{equation}
where $T_{AB}$ is the energy-momentum tensor and
\begin{equation}
G^{\;\; A}_{(p)\;\,B}=
-\frac{1}{2^{p+1}}
\delta^{A i_1 \ldots i_{2p}}_{Bj_1\ldots j_{2p}} 
R^{j_1j_2}_{\phantom{j_1j_2}i_1i_2}
\ldots
R^{j_{2p-1}j_{2p}}_{\phantom{j_{2p-1}j_{2p}}i_{2p-1}i_{2p}}.
\end{equation}

We have seen that branes with codimension 1 or 2 admit regular
solutions in Einstein gravity~\footnote{We consider pure tension
  branes now for 
the case of codimension 2, the subtleties associated with a general
codimension 2 energy-momentum tensor will be discussed below.} whereas
branes of codimension 3 or higher do not. 
The reason is that the Riemann tensor of a general
non-singular metric can only have uni- and bi-dimensional delta-like
behaviour through discontinuities of the first derivatives of the
metric but not higher dimensional ones. A natural way of obtaining
higher codimension branes is then
by considering two or more branes that intersect 
and higher orders in the curvature expansion in such a way
that the product of Riemann tensors (or contractions thereof) gives
the product of delta functions at the intersection of the branes. 
The use of the Lanczos-Lovelock Lagrangian ensures that 
unaccountable for singularities of the type $\delta(y)^2$ are absent
in the higher curvature corrections. In this   
section we will illustrate this mechanism for the generation of 
higher dimensional delta functions in the Einstein-Lanczos-Lovelock
tensor, Eq.(\ref{Lovelock:EoM}), with the simplest examples, 
namely the intersection of two codimension 1 branes, a codimension 1
with a codimension 2 brane and two codimension 2 branes, giving rise
to, respectively, codimension 2, 3 and 4 branes living in the
intersection. We will then 
generalise this construction to the intersection of an arbitrary
number of codimension 1 or 2 branes. In section~\ref{examples:sect} 
we give two explicit examples of solutions, first a codimension 3 brane arising at the
intersection of a codimension 1 brane with a codimension 2 brane in
an $AdS_5\times S^2$ background in Gauss-Bonnet gravity 
and second a string motivated 10D solution where 5-branes and 3-branes are obtained at the intersections of,
respectively, two and three codimension 2 branes in a
(Minkowski)$_4\times S^6$ background.

\subsection{Intersection of codimension 1 branes}

In this section we review the example of the intersection of two
codimension 1 branes, that has been recently used as a way to obtain a
codimension 2 brane world living in the intersection
\cite{Lee:2004hh}. The backreaction on the background for a
codimension 1 brane can be dealt with assuming discontinuities in the
first derivatives of the metric with respect to 
the orthogonal coordinate at the position of the brane, that produces 
a jump in the extrinsic curvature as one goes from one side of the
brane to the other \cite{Israel:rt}. In the intersection of two such
branes a two-dimensional delta function is generated by the second
Lovelock (Gauss-Bonnet) term as follows.  
We consider six-dimensional space-time and take following ansatz for
the metric,  
\begin{equation}
\mathrm{d}s^2=
g_{\mu\nu}(x,y,z)\, \mathrm{d}x^\mu \mathrm{d}x^\nu
-W^2(x,y,z)\, \mathrm{d}y^2 -L^2(x,y,z) \,\mathrm{d}z^2.
\end{equation}
We also consider a $Z_2$ symmetry for each of the ``extra'' dimensions
in $z=0$ and $y=0$, where we locate two codimension 1 branes spanning,
respectively, the $(x^\mu,y)$ and $(x^\mu,z)$ coordinates. The
relevant components of the Riemann tensor for the matching conditions
(those with a delta like behaviour) are 
\begin{eqnarray}
R^{y\mu}_{\phantom{y\mu}y\nu}&=& \frac{1}{2}
\frac{g^{\mu\rho}\ddot{g}_{\rho\nu}}{W^2}+\ldots, \\
R^{z\mu}_{\phantom{z\mu}z\nu}&=& \frac{1}{2}
\frac{g^{\mu\rho}g^{\prime\prime}_{\rho\nu}}{L^2}+\ldots, \\
R^{yz}_{\phantom{yz}yz}&=& 
\frac{\ddot{L}}{W^2L}+\frac{W^{\prime\prime}}{L^2W}+\ldots,
\end{eqnarray}
where a dot and a prime denote, respectively, derivatives with respect
to $y$ and $z$ and we have used the fact that
for a discontinuous first derivative we have
\begin{equation}
\ddot{g}_{\mu\nu}(x,y,z)=2 \dot{g}_{\mu\nu}(x,0^+,z)\; \delta(y)+ \dots,
\end{equation}
and similarly with the other components. 
Using the general expression for the different terms
in the Lovelock tensor, Eq.(\ref{Lovelock:EoM}), we obtain the
following singular terms in the Einstein tensor
(Lovelock term of order 1)
\begin{eqnarray}
G^{(1)}_{\mu\nu}&=&
\frac{1}{W^2}
\left[-g_{\mu\nu}\left( 2 \frac{\dot{L}}{L}+
  g^{\rho\sigma}\dot{g}_{\rho\sigma} \right) + \dot{g}_{\mu\nu}
  \right] \delta(y)
\nonumber \\
&+&\frac{1}{L^2}
\left[-g_{\mu\nu}\left( 2 \frac{W^\prime}{W}+
  g^{\rho\sigma}g^\prime_{\rho\sigma} \right) + g^\prime_{\mu\nu}
  \right] \delta(z) + \dots, \label{co11}
\\
G^{(1)}_{yy}&=&
\frac{W^2}{L^2} g^{\rho \sigma} g^\prime_{\rho\sigma} \delta(z)+ \dots,\label{co12} \\
G^{(1)}_{zz}&=&
\frac{L^2}{W^2} g^{\rho \sigma} \dot{g}_{\rho\sigma} \delta(y)+ \dots \label{co13},
\end{eqnarray}
where the dots represent terms without delta functions.
As expected there is a term proportional to $\delta(y)$ in the
$(\mu\nu)$ and $(zz)$ components of the Riemann tensor and a term
proportional to $\delta(z)$ in the $(\mu\nu)$ and $(yy)$ ones,
and therefore one can find solutions that represent two codimension 1 branes
that intersect at the points $y=z=0$. In this intersection, when including the second Lovelock (Gauss-Bonnet) term we have to consider the presence of a codimension 2 brane with non-zero tension. This can de seen by computing the coefficient of $\delta(y)\delta(z)$ appearing in $G^{(2)}_{MN}$,
\begin{eqnarray}
G^{(2)}_{\mu\nu}&=&
-\frac{4}{W^2L^2} 
\bigg[
g_{\mu\nu}\Big( 
g^{\rho\sigma} \dot{g}_{\rho\sigma}
g^{\lambda \tau} g^\prime_{\lambda\tau}
-g^{\lambda\sigma} \dot{g}_{\rho\tau}
g^{\tau \rho} g^\prime_{\rho\lambda}
\Big)
\nonumber \\
&&\quad\quad
-\dot{g}_{\mu\nu} g^{\rho\sigma}g^\prime_{\rho\sigma}
- g^\prime_{\mu\nu} g^{\rho\sigma} \dot{g}_{\rho\sigma}
+ \dot{g}_{\mu\rho} g^{\rho\sigma} g^\prime_{\sigma \nu}
+ g^\prime_{\mu\rho} g^{\rho\sigma} \dot{g}_{\sigma\nu}
\bigg]\, \delta(y) \delta(z) +\dots
\end{eqnarray}
Now the dots represent terms without delta functions or with just one
delta. Notice that the two-dimensional delta function only appears
along the $(\mu,\nu)$ coordinates, the coordinates of a 3-brane
sitting in $z=y=0$. The terms with just one delta appearing in
$G^{(2)}_{MN}$ modify the matching conditions for the codimension 1
branes, Eqs.(\ref{co11}$-$\ref{co13}), but we do not write
them explicitly since they are somewhat complicated and do not
contribute anything to the discussion (general expressions for the
matching conditions of codimension 1 branes in Einstein-Gauss-Bonnet
gravity can be found in \cite{Maeda:2003vq}).

\subsection{Intersection of codimension 1 and codimension 2 branes}

In this section we shall describe the intersection of a codimension 1
brane with a codimension 2 one, and we will see that, when including
the Lovelock terms, one needs the presence of a non-trivial
energy-momentum density localized in the intersection, leading to the
first example of a 
codimension 3 brane with a non-singular induced metric. In order to do that
consider a seven-dimensional space time with the following metric
\begin{equation}
\mathrm{d}s^2=
g_{\mu\nu}(x,r,y)\, \mathrm{d}x^\mu \mathrm{d}x^\nu
-W^2(x,y)\, \mathrm{d}r^2 -L^2(r,x,y) \,\mathrm{d}\theta^2-\mathrm{d}y^2,
\end{equation}
where as in the previous section we consider a $Z_2$ symmetry
around $y=0$, $\theta$ has period $2\pi$ and in order for these coordinates to represent a
codimension 2 submanifold at $r=0$ we must have $L\simeq \beta
r+\mathcal{O}(r^2)$ for small $r$. Similarly to the case of the
codimension 1 brane in which the jump in the normal derivatives
generate a one-dimensional delta function, if the slope of the
function $L$ is not one in $r=0$ ($\beta \neq 1$), a conical
singularity (\textit{i.e.} a two-dimensional delta function)
is generated in the Einstein tensor at that point 
\begin{equation}
\frac{\partial_r^2 L}{L}=-(1-\beta) \frac{\delta(r)}{L} +\ldots
\end{equation}
We will take this as the only source of $\delta(r)$ behaviour. The possibility of considering $\partial_rg_{\mu \nu}|_{r=0^+} \neq 0$, so $\partial^2_rg_{\mu \nu} \sim \delta(r)$ would generate also a two-dimensional delta function in the Gauss-Bonnet term \cite{Bostock:2003cv}, but it would lead to a divergent Ricci tensor as one approaches $r=0$ since
\begin{eqnarray}
R_{\mu\nu}= \frac{1}{2} \frac{L^\prime}{L} \partial_r g_{\mu\nu}
+ \ldots = \frac{\partial_r g_{\mu\nu}}{2r} + \mathcal{O}(1)
\end{eqnarray}
near the brane, so we will consider only solutions in which  $\partial_rg_{\mu \nu}|_{r=0^+}=0$.
The relevant (distributional) components of the Riemann tensor are then
\begin{eqnarray}
R^{y\mu}_{\phantom{y\mu}y\nu}&=& \frac{1}{2}
\frac{g^{\mu\rho}\ddot{g}_{\rho\nu}}{W^2}+\ldots, \\
R^{yr}_{\phantom{yr}yr}&=& 
\frac{\ddot{W}}{W}+\ldots,\\
R^{y\theta}_{\phantom{y\theta}y\theta}&=& 
\frac{\ddot{L}}{L}+\ldots,\\
R^{r\theta}_{\phantom{r\theta}r\theta}&=& 
\frac{\partial_r^2L}{W^2L}+\ldots,
\end{eqnarray}
where a dot denotes a derivative with respect to $y$ and we have, as before,
\begin{equation}
\ddot{g}_{\mu\nu}(x,r,y)=2 \dot{g}_{\mu\nu}(x,r,0^+)\; \delta(y)+ \dots,
\end{equation}
and similarly with the other components.
Again it is straightforward to obtain the delta-like components of the Einstein tensor:
\begin{eqnarray}
G^{(1)}_{\mu\nu}&=&
(1-\beta) g_{\mu\nu}\frac{1}{W^2} \frac{\delta(r)}{L}
+\left[-g_{\mu\nu}\left( 2 \frac{\dot{L}}{L}+ 2 \frac{\dot{W}}{W}+
  g^{\rho\sigma}\dot{g}_{\rho\sigma} \right) 
+ \dot{g}_{\mu\nu}
  \right] \delta(y)+\dots,\label{cod1-cod2:E:munu}
\\
G^{(1)}_{yy}&=&
-(1-\beta) \frac{1}{W^2} \frac{\delta(r)}{L} 
\label{cod1-cod2:E:yy}+\dots, \\
G^{(1)}_{rr}&=&
W^2
\Big[g^{\rho\sigma}\dot{g}_{\rho\sigma}+2\frac{\dot{L}}{L} \Big]
\delta(y)+\dots, \\
G^{(1)}_{\theta\theta}&=&
L^2
\Big[g^{\rho\sigma}\dot{g}_{\rho\sigma}+2\frac{\dot{W}}{W} \Big]
\delta(y)+\dots, 
\end{eqnarray}
where the dots represent terms without deltas while the three-dimensional delta function appearing in the Gauss-Bonnet term is given by 
\begin{eqnarray}
G^{(2)}_{\mu\nu}&=&
(1-\beta)\frac{1}{W^2} 
\bigg[
g_{\mu\nu} g^{\rho\sigma}\dot{g}_{\rho\sigma} - \dot{g}_{\mu\nu}
\bigg] \delta(y) \frac{\delta(r)}{L}+\dots
\label{cod1-cod2:GB:munu}
\end{eqnarray}
where the dots represent terms without three-dimensional delta
functions. Notice however that in these terms that we have not written
there are contributions proportional to $\delta(y)$ and $\delta(r)/L$
that will modify the matching conditions obtained from the Einstein
part. These corrections can be seen as small for the codimension 1
brane (we do not write them here for the same reasons as in the
previous section), but as we have previously explained they are
crucial for the codimension 2 brane, since in the Einstein term the
delta-like contribution is proportional to the brane induced metric
[see Eqs.(\ref{cod1-cod2:E:munu},\ref{cod1-cod2:E:yy})] so in order to
find solutions for a codimension 2 brane that has a general
energy-monentum tensor (not just pure tension) on has to consider the
contribution of the Gauss-Bonnet term to the matching
condition\footnote{This property,  
comes from a term in Eq.(\ref{cod1-cod2:GB:munu}) 
proportional to
$\hat{G}^{(1)}_{mn} \frac{\delta(r)}{L}$ that we have not
explicitely written, where $\hat{G}^{(1)}_{mn}$ represents Einstein
tensor for the induced metric on the codimension 2 brane.}.

Notice that the matching condition for the codimension 3 brane at the
intersection inherits the richer structure of its codimension 1
parent. (The matching condition is indeed identical to that of a
codimension 1 brane in five dimensions.) It is worth pointing out that in 7 dimensions we have also the third Lovelock term at our disposal. This term would contribute to all the matching conditions, but the structure of the Einstein-Lanczos-Lovelock tensor, Eq.(\ref{Lovelock:EoM}), ensures that terms proportional to $\delta(y)^2$ or $\delta(r)^2$ will not appear. Also, it is easy to see that terms proportional to $\delta(y)$ will only appear along the $(\mu, \nu)$, $(r,r)$ and $(\theta,\theta)$ components, those proportional to $\delta(r)/L$ will only appear along the  $(\mu, \nu)$ and  $(y,y)$ components, while those proportional to $\delta(y)\delta(r)/L$ will only appear in the $(\mu, \nu)$ components, just contributing subleading corrections to the matching conditions already obtained.

\subsection{Intersection of 2 codimension 2 branes}
 
We now turn to the intersection of two codimension 2 branes and we will see that, when including higher Lovelock terms in the action, one needs a codimension 4 brane living in the intersection in order to satisfy the matching conditions. We consider eight space time
dimensions with the following ansatz for the metric,
\begin{equation}
\mathrm{d}s^2=g_{\mu\nu}(x,r_1,r_2)\, \mathrm{d}x^\mu \,
\mathrm{d}x^\nu 
- \mathrm{d}r_1^2\,- L_1^2(r_1)
 \, \mathrm{d}\theta_1^2 
- \mathrm{d}r_2^2\,- L_2^2(r_2)
 \, \mathrm{d}\theta_2^2,
\end{equation}
where as usual $\theta_i$ have period $2\pi$, $L_i=\beta_i r_i+\mathcal{O}(r_i^2)$, $i=1,2$ and, in order
to avoid curvature singularities as we approach the branes we consider
$\partial_{r_i} g_{\mu\nu}|_{r_i=0^+}=0$.
According to our experience with codimension 2 branes and
intersections between branes one can expect that the Einstein tensor 
will allow us to find solutions for a pure tension codimension 2 brane, including the Gauss-Bonnet
term we will be able to satisfy the matching conditions for general codimension 2 brane but only for a pure tension
codimension 4 brane at the intersection whereas the third Lovelock
term, that is also available in eight dimensions, 
will give enough freedom to match a general energy-momentum
tensor for the codimension 4 brane. This intuition is in fact correct
as we show now. 


Since we have assumed that $\partial_{r_i} g_{\mu\nu}|_{r_i=0^+}=0$ the only source of delta functions in the Riemann tensor are the conical singularities, and the relevant components of this tensor are then
\begin{eqnarray}
R^{r_i\theta_i}_{\phantom{r_i\theta_i}r_i\theta_i}=
\frac{\partial_{r_i}^2 L_i}{L_i},
\end{eqnarray}
where again $i=1,2$. 
The Einstein tensor has the following singular components
\begin{eqnarray}
G^{(1)}_{\mu\nu}&=&
-g_{\mu\nu} \sum_{i=1}^2
\Big[(1-\beta_i) \frac{\delta(r_i)}{L_i}\Big]+\dots,
\\
G^{(1)}_{r_ir_i}&=&(1-\beta_j) \frac{\delta(r_j)}{L_j}+\dots,\quad j\neq i,
\\
G^{(1)}_{\theta_i \theta_i}&=&(1-\beta_j)L_j^2
\frac{\delta(r_j)}{L_j}+\dots,\quad j\neq i, 
\end{eqnarray}
while the Gauss-Bonnet tensor has the following codimension 4 singularity,
\begin{eqnarray}
G^{(2)}_{\mu\nu}&=&
-4 g_{\mu\nu} (1-\beta_1)(1-\beta_2) 
\frac{\delta(r_1)}{L_1}
\frac{\delta(r_2)}{L_2}+\dots,\label{c4}
\end{eqnarray}
plus codimension 2 singularities that we do not explicitely write. 
These codimension 2 deltas generated in the Gauss-Bonnet term are
proportional to the Einstein tensors for the induced metrics on the 
corresponding 5-branes. (The $(r_i,r_i)$ and 
$(\theta_i, \theta_i)$ components will of course have the
corresponding contribution proportional to $\delta(r_{j\neq i})/L_j$.) 
Notice that, as expected, if we do not include the next Lovelock term,
we will be able to find solutions only for a pure tension 3-brane,
since the $\delta^{(4)}$ term is proportional to the brane induced
metric [Eq.(\ref{c4})]. In 
order to be able to satisfy the matching conditions for a general
energy-momentum tensor for the 3-brane at 
the intersection we have to make use of the third Lovelock term that
has the following codimension 4 
delta function
\begin{eqnarray}
G^{(3)}_{\mu\nu}=24(1-\beta_1)(1-\beta_2) \hat{G}^{(1)}_{\mu\nu}(g) 
\frac{\delta(r_1)}{L_1} \frac{\delta(r_2)}{L_2}+\dots,
\end{eqnarray}
where $\hat{G}^{(1)}_{\mu\nu}(g)$ is the Einstein tensor for the induced
metric on the 3-brane. As always there are lower codimension delta
functions in this tensor not explicitly written that will modify the
matching conditions for the codimension 2 branes. In fact is easy to
see that these corrections will take the form of the Gauss-Bonnet
(second Lovelock term) for the induced metric of the respective
5-branes and as we will see in the next subsection this structure is
straightforwardly generalisable to higher (co)dimensions. 
  
\subsection{General intersection of codimension 1 and codimension 2
  branes} 

We have shown how Lanczos-Lovelock gravity allows us to obtain
up to codimension 4 delta functions in the generalized Einstein tensor
in an otherwise regular background at the intersection of two branes
of codimensions 1 or 2. These solutions require then the presence of a
higher codimension branes living in these intersections. 
In this section we are going to generalise these examples by discussing 
the intersection of an arbitrary number of codimension 1 and 2 branes.

In a $D$-dimensional space time we have non-trivial Lovelock terms up
to order $p_\mathrm{max}=\left[\frac{D-1}{2}\right ]$, where as usual
the square brackets denote integer part. According to our previous
discussion, the fact that we can get one codimension 1 or 2 delta functions in
the Riemann tensor of a regular metric (and its contractions) 
implies that up to $p_\mathrm{max}$ branes can have a non-trivial
intersection. Consider the intersection of $m_1$ codimension 1 branes
and $m_2$ codimension 2 branes, where $m_1+m_2\leq
p_\mathrm{max}$. The brane at the common intersection has codimension 
$m_1+2m_2$, which can be up to $2 p_\mathrm{max}$ when all the branes
we use are codimension 2. This means that in $D$ dimensions we can
in principle match up to $0-$branes or $1-$branes for $D$ odd and
even, respectively. 
We have obtained those numbers by just counting
the number of powers of the Riemann tensor we have. From the general
form of the Lovelock equations of motion and the examples we have
discussed above it is however evident 
that for an arbitrary $D$ (we have to
consider $D\geq 5$ if we want to have any nontrivial intersection
although the argument carries on for lower dimensions with just one
brane) solutions representing all the lower dimensional branes down to $0-$ or $1-$ branes
can indeed be obtained by means of brane intersections. 

 When we considered the intersection of two codimension 2 branes, we had to consider the contribution of the third
Lovelock term to the matching conditions to be able to satisfy them for a general 4-brane energy-momentum tensor. Remarkably, we found the interesting result that the matching condition implies that the induced metric at this intersection satisfies the lower dimensional Einstein equations, analogously to what happens for the codimension 2 brane.  In fact it is easy to see that the equations of motion (obtained from the matching conditions) for the induced metric
on the world-volume of a codimension 2 brane in Lovelock gravity of
order $p$ correspond to the Lovelock
equations of order $p-1$. So for a D-dimensional metric of the form
\begin{equation}
\mathrm{d}s^2=g_{\mu\nu}(x,r)\, \mathrm{d}x^\mu \,
\mathrm{d}x^\nu 
- \mathrm{d}r^2\,- L^2(r)
 \, \mathrm{d}\theta^2,
\end{equation}
where as always $0\leq \theta \leq 2 \pi$, $L=\beta r+\mathcal{O}(r^2)$ and $\partial_r g_{\mu\nu}|_{r=0^+}=0$, the Lovelock tensor of order $p$ will contribute to the matching
condition
\begin{eqnarray}
G^{\;\; \mu}_{(p)\;\,\nu}|_{\mathrm{Cod\,2}}&=&
-\frac{4p}{p+1}
\delta^{\mu i_1 \ldots i_{2p-2}}_{\nu j_1\ldots j_{2p-2}} 
R^{j_1j_2}_{\phantom{j_1j_2}i_1i_2}
\ldots
R^{j_{2p-3}j_{2p-2}}_{\phantom{j_{2p-3}j_{2p-2}}i_{2p-3}i_{2p-2}}
R^{r\theta}_{\phantom{r\theta}r\theta}+\dots
\nonumber \\
&=&
-2p(1-\beta)\frac{\delta(r)}{L}\hat{G}^{\;\;\mu}_{(p-1)\;\,\nu}+\dots
\end{eqnarray}
In the last
equality the hat represents the corresponding Lovelock tensor computed
with the induced metric on the brane [$i.e.$ $g_{\mu\nu}(x,0)$]. 
Similarly, if we consider the intersection of $m$ codimension 2
branes with a metric of the type
\begin{equation}
\mathrm{d}s^2=g_{\mu\nu}(x,r_i)\, \mathrm{d}x^\mu \,
\mathrm{d}x^\nu 
-\sum_{i=1}^m (\mathrm{d}r_i^2\,+ L_i^2(r_i)
 \, \mathrm{d}\theta_i^2),
\end{equation}
with $0\leq \theta_i \leq 2 \pi$, $L_i=\beta_i r_i+\mathcal{O}(r_i^2)$ and $\partial_{r_i} g_{\mu\nu}|_{r_i=0^+}=0$, the contribution of the order $p$ Lovelock term to the matching condition of the
intersection (at all $r_i=0$) reads
\begin{equation}
G^{\;\; \mu}_{(p)\;\,\nu}|_{\mathrm{m \,Cod\, 2}}=
(-1)^m \frac{2^m p!}{(p-m)!} \hat{G}^{\;\;\mu}_{(p-m)\;\,\nu} \prod_{i=1}^m
(1-\beta_i) 
\frac{\delta(r_i)}{L_i}+\dots, 
\end{equation}
where $\hat{G}^{\;\;\mu}_{(p-m)\;\,\nu}$ is the $(p-m)$-th order Lovelock tensor for the induced metric. In particular
if we restrict ourselves to an even number of dimensions (in order to
eventually obtain 3-branes), $D=2n$, we have at our disposal $n-1$
Lovelock terms. The order $n-2$ Lovelock term allows us to match a
pure Einstein 3-brane at the intersection of $n-2$ codimension 2
branes, 
\begin{equation}
G^{\;\; \mu}_{(n-2)\;\,\nu}|_{\mathrm{(n-2) \,Cod\, 2}}=
(-2)^{n-3} (n-2)! \delta^{\mu}_{\nu} \prod_{i=1}^{n-2}
(1-\beta_i) 
\frac{\delta(r_i)}{L_i}+\dots, 
\end{equation}
 whereas using the highest, order $n-1$, Lovelock term we can match a
 general energy-momentum tensor on the 3-brane at the intersection of
 the $n-2$ codimension 2 branes since
\begin{equation}
G^{\;\; \mu}_{(n-1)\;\,\nu}|_{\mathrm{(n-2) \,Cod\, 2}}=
(-2)^{(n-2)}  (n-1)! \hat{G}^{\;\;\mu}_{(1)\;\,\nu} \prod_{i=1}^{n-2}
(1-\beta_i) 
\frac{\delta(r_i)}{L_i}+\dots
\end{equation}
and we obtain the Einstein equation for the induced metric on the
brane. Thus, the structure of the matching conditions for the
intersection of an arbitrary number of codimension 2 branes in
Lovelock gravity has a ``russian doll'' structure, with the induced
metric in the worldvolume of each brane satisfying the Lovelock
equations that corresponds with its dimensionality. 
An example corresponding to $D=10$ will be worked out in the next section.

Once we have written the general expression for the intersection of an
arbitrary number of codimension 2 branes, we can use it to study the
intersection of $m_1$ codimension 1 and $m_2$ codimension 2
branes in two steps. First consider the intersection of the $m_2$
codimension 2 branes. The resulting matching conditions for the
brane at such intersection in order $p$ Lovelock gravity are, as
we have just shown, the order $p-m_2$ Lovelock equations for the
induced metric on the brane intersection. We are therefore left with
the problem of what the equations of motion are for the metric induced
at the intersection of an arbitrary number of codimension 1
branes. The solution to that problem cannot be written in as neat a
way as in the case of codimension 2 branes. The reason is that in the
codimension 2 case we are assuming that the extrinsic curvature is
zero ($i.e.$ $\partial_r g_{\mu\nu}|_{r=0^+}=0$, with
$g_{\mu\nu}|_{r=0}$ the brane induced metric) and we are left with
some equations for the induced metric that have a closed form, and do
not depend on the bulk structure. For the codimension 1 case this
cannot be done since the discontinuity in the extrinsic curvature is
the 
only source of delta functions, and to obtain the equations that
relate the induced metric with the brane energy-momentum tensor one
has to write down the Einstein tensor for the induced metric in terms
of the extrinsic curvature plus corrections (this can be done using
the Gauss-Codazzi formalism), and the equations of motion for the induced metric can not be obtained in
a closed form \cite{Shiromizu:1999wj}. 

\section{Explicit examples \label{examples:sect}}

So far we have only computed matching conditions in Lovelock gravity
and we have seen that higher Lovelock terms 
give enough freedom to match higher codimension branes at brane
intersections. This is an important result, being the first example of
local solutions for higher codimension branes in a regular
background. It still remains the question however of whether such
solutions can be consistent globally. In this section we
demonstrate with a couple of simple, but phenomenologically relevant
examples that global solutions for higher codimension branes can
indeed be obtained. In Ref.~\cite{Lee:2004hh} global solutions are
found for a codimension 2 brane arising at the intersection of two
codimension 1 branes in Einstein-Gauss-Bonnet gravity. Here we will
give two examples of global solutions with 
one or two codimension 3 branes arising at the intersection
of a codimension 1 and a codimension 2 brane (in an $AdS_5\times S^2$
background) and two codimension 6 (codimension 4) branes at the
intersection of three (two) codimension 2 branes in a
(Minkowski)$_4\times S^6$ background. The two solutions shown here
do not present any horizon and the background is regular
everywhere. We have not made any assuption about the sign of the brane
tensions. This could be relevant for questions of 
stability~\cite{Charmousis:2003sq} that are beyond the scope of the
present study.

\subsection{A 7D model}

A simple example for a codimension 3 brane arising at the intersection
of a codimension 2 with a codimension 1 brane in a regular globally
defined background can be obtained in $AdS_5\times S^2$ with the
following metric
\begin{equation}
\mathrm{d}s^2 \, =
\mathrm{e}^{-k|y|} \eta_{\mu\nu} \mathrm{d}x^\mu\, \mathrm{d}x^\nu\,
- \mathrm{d}y^2\,- R^2(\mathrm{d}\theta^2\, + \beta^2 \sin^2\theta\,
\mathrm{d}\phi^2\,),
\end{equation}
where we have imposed a $Z_2$
symmetry around $y=0$, $\theta$ ranges from $0$ to $\pi$ while $\phi$ has the standard periodicity of $2\pi$ and we allowed for an arbitrary deficit angle
$1-\beta$. It is easy to see that this metric is a solution of
Einstein-Gauss-Bonnet equations in the bulk with the following
energy-momentum tensor 
\begin{equation}
T^\mathrm{Bulk}_{MN}=
\left(\begin{array}{ccc}
\mathrm{e}^{-k|y|} \eta_{\mu\nu} \Lambda_1 & & \\
& - \Lambda_1 & \\
& & \kappa_{ij} \Lambda_2
\end{array}
\right),
\end{equation}
where $\kappa_{ij}$ is the metric on the sphere and the two constants
$k$ and $R$ are related to the ones appearing in the energy-momentum tensor as
\begin{eqnarray}
\Lambda_1&=&
-\frac{3}{4}k^2(2+k^2 \alpha_2) + \frac{1+6k^2\alpha_2}{R^2}, \\
\Lambda_2&=&
-\frac{5}{4}k^2(2+3k^2 \alpha_2).
\end{eqnarray}
Such an inhomogeneous $vev$ for the energy-momentum tensor can be obtained through the flux of a 2- or 5-form (Freund-Rubin compactification \cite{Freund:1980xh}) or through an anisotropic Casimir effect \cite{Candelas:ae}, for instance.

The jump in the extrinsic curvature at $y=0$ gives rise to a delta function contribution to the Einstein-Gauss-Bonnet tensor that can be interpreted as the backreaction due to a brane located at $y=0$ with the following energy-momentum tensor
\begin{equation}
T^{\mathrm{Cod}\,1}_{MN}=
\left (
\begin{array}{ccc}
3k(1+k^2\alpha_2)\mathrm \eta_{\mu\nu} \delta(y) && \\
&0& \\
&& 4k(1+3k^2\alpha_2)k_{ij} \delta(y)
\end{array}
\right).
\end{equation}
Notice that we need an inhomogeneous form for the brane tension. This
can be easily generated again by considering the magnetic flux of a
$U(1)$ gauge field or the flux of a 4 form localized on the brane. In
any case we find that these values have to be fine tuned with respect
to the bulk energy-momentum tensor, as in the original Randall-Sundrum
model \cite{Randall:1999ee}\footnote{This is a generic feature of
  codimension 1 models. Since the bulk geometry determines the jump in
  the extrinsic curvature one has to fine tune the brane tension with
  respect to bulk parameters. There have been however attempts to get
  rid of this fine tuning by coupling the brane to a scalar field that
  were the origin of the so-called selftuning models
  \cite{Arkani-Hamed:2000eg}.}. 
A value of $\beta \neq 1$ can be interpreted as being the backreaction
induced by two codimension 2 branes at 
$\theta=0,\pi$ (we can get rid of one of these branes by considering a
$Z_2$ orbifolding of the sphere with respect to the equatorial plane)
with energy-momentum tensor given by 
\begin{equation}
T^{\mathrm{Cod}\,2}_{MN}=
\left (
\begin{array}{ccc}
(1-\beta)(1+6k^2\alpha_2)\mathrm{e}^{-k|y|} \eta_{\mu\nu}
\frac{\delta(\sin\theta)}{R^2\beta \sin \theta} && \\
&
-(1-\beta)(1+6k^2\alpha_2)
\frac{\delta(\sin\theta)}{R^2\beta \sin \theta}
& 
\\
&& 0
\end{array}
\right).
\end{equation}
Now the brane tension can take any value (since $\beta$ is a free parameter of the solution) without modifying the brane geometry. This is a reflection of the fact that in codimension 2 the brane tension does not have to be fine tuned with respect to bulk parameters and thus provide an interesting starting point for building selftuning models \cite{Chen:2000at,Carroll:2003db}.
Finally, the Gauss-Bonnet term has a $\delta^{(3)}$ contribution that gives the matching of the codimension 3 brane at the intersection with energy-momentum tensor
\begin{equation}
T^{\mathrm{Cod}\,3}_{MN}=
\left (
\begin{array}{ccc}
-12\alpha_2 k(1-\beta)\mathrm \eta_{\mu\nu}
\delta(y)\frac{\delta(\sin\theta)}{R^2\beta \sin \theta} && \\
&0&
\\
&& 0
\end{array}
\right).
\end{equation}
The tension of this brane is fixed in terms of the other parameters of the solution.

\subsection{A 10D model}

We finally want to show an explicit example of the intersection
of codimension 2 branes. Instead of describing the simplest case of
two codimension 2 branes we consider the string motivated one 
of a codimension 
6 brane arising at the intersection of three codimension 2 branes in
ten-dimensional space time. In particular we take our space time to be
(Minkoski)$_4\times S^6$ with the following metric
\begin{eqnarray}
\mathrm{d}s^2&=&
\eta_{\mu\nu} \mathrm{d}x^\mu\, \mathrm{d}x^\nu\,
- R^2
\Big[\mathrm{d}\theta_1^2\,+\beta_1^2\sin^2\theta_1 \mathrm{d}\phi_1^2\,
\nonumber \\
&&+\cos^2\theta_1
\Big(\mathrm{d}\theta_2^2\,+\beta_2^2\sin^2\theta_2 \mathrm{d}\phi_2^2\,
+\cos^2\theta_2
(\mathrm{d}\theta_3^2\,+\beta_3^2\sin^2\theta_3 \mathrm{d}\phi_3^2\,
) \Big)
\Big],
\end{eqnarray} 
where $0\leq\theta_{1,2}\leq \pi/2$, $0\leq\theta_{3}\leq \pi$, 
$0\leq \phi_i \leq 2 \pi$ and 
the three arbitrary constants $\beta_1$, $\beta_2$
and $\beta_3$ that are allowed by the symmetries of the six-sphere will
allow us to match the energy-momentum tensor of three codimension 2
branes located each one at $\sin\theta_i=0$, for $i=1,2,3$.
Even though up to the fourth order Lovelock term is available in ten
dimensions, in 
our case due to the factorizable form of the metric and the
fact that the four non-compact dimensions are flat, all the components
of the Lovelock term of order four vanish. Therefore we will show that
this metric represents a well behaved 
globally defined solution with three codimension
2 branes intersecting by pairs on three codimension 4 branes and the three
of them at two codimension 6 branes in Lovelock gravity\footnote{The
  submanifold defined by the condition $\sin\theta_i=0$ is nothing but
  (Minkowski)$_4\times S^4$ (notice that $\theta_3=0$ covers one
  hemisphere and we need $\theta_3=\pi$ to cover the full $S^4$)  
while one can check that imposing the
  conditions $\sin\theta_2=\sin\theta_3=0$ we are left with a
  submanifold that corresponds to (Minkowski)$_4\times S^2$ (and the
  same for every other pair of theta angles). Finally, the condition
  $\sin\theta_1=\sin\theta_2=\sin\theta_3=0$ results on two 3-branes
  [(Minkowski)$_4\times S^0$].}. It is indeed easy to
see that the bulk equations of motion are 
satisfied for the following energy-momentum tensor 
\begin{equation}
T_{MN}^\mathrm{Bulk}=
\left(
\begin{array}{cc}
\eta_{\mu\nu} \Lambda_1 & \\
& \kappa_{ij} \Lambda_2
\end{array}
\right),
\end{equation}
where $\kappa_{ij}$ is the metric for the 6-sphere and the constants
$\Lambda_{1,2}$ have to be fine-tuned to get four-dimensional flat
space, being related to $R$ as
\begin{eqnarray}
\Lambda_1&=& \frac{15}{R^2}\left(1-\frac{12}{R^2}\alpha_2 +
\frac{24}{R^4}\alpha_3 \right), \\
\Lambda_2&=& \frac{10}{R^2} \left( 1-\frac{6}{R^2} \alpha_2 \right).
\end{eqnarray} 
The asymmetric energy-momentum tensor can be obtained as before using
the \textit{vev} of the corresponding form or through anisotropic
Casimir effect. 

Now we turn to the branes. A non-trivial value of the deficit angles,
$\beta_i\neq 1$, induces a conical singularity at $\sin\theta_i=0$,
matching an energy-momentum for a codimension 2 brane with the
following non-vanishing components
\begin{eqnarray}
T^{\mathrm{Cod}\,2}_{\mu\nu}(i)&=&
\eta_{\mu\nu}(1-\beta_i)\left(1-\frac{24}{R^2}
\alpha_2+\frac{72}{R^4}\alpha_3\right)
\frac{\delta(\sin\theta_i)}{\sqrt{\kappa_i}},  \\
T^{\mathrm{Cod}\,2}_{kl}(i)&=&
\kappa_{kl}(1-\beta_i)\left(1-\frac{12}{R^2}
\alpha_2\right)
\frac{\delta(\sin\theta_i)}{\sqrt{\kappa_i}}, \quad k,l\neq \theta_i,\phi_i,
\end{eqnarray}
where $i=1,2,3$ and $\kappa_i$ is the determinant of the $2\times 2$
submatrix of the sphere metric corresponding to the coordinates
$\theta_i,\phi_i$. Note that, according to our general discussion,
the contribution of each Lovelock term to the matching condition for
one codimension 2 brane is proportional to the previous Lovelock term
for the induced metric on the 
brane. The fact that we are forcing the worldvolume of the brane to be
(Minkowski)$_4\times S^4$ implies that the brane energy-momentum
tensor has to be again inhomogeneous (we could consider magnetic
fluxes for gauge fields localized on the brane in order to generate
it, for instance).  

These 7-branes intersect by pairs on codimension 4 branes with the
following energy-momentum tensor, 
\begin{eqnarray}
T^{\mathrm{Cod}\,4}_{\mu\nu}(1)&=&
-\eta_{\mu\nu}(1-\beta_2)(1-\beta_3)\left(4\alpha_2-\frac{24}{R^2}
\alpha_3\right)
\frac{\delta(\sin\theta_2)}{\sqrt{\kappa_2}}
\frac{\delta(\sin\theta_3)}{\sqrt{\kappa_3}},  \\
T^{\mathrm{Cod}\,4}_{ij}(1)&=&
-4\alpha_2\kappa_{ij}(1-\beta_2)(1-\beta_3)
\frac{\delta(\sin\theta_2)}{\sqrt{\kappa_2}}
\frac{\delta(\sin\theta_3)}{\sqrt{\kappa_3}},\quad i,j=\theta_1,\phi_1,  
\end{eqnarray}
where we have shown as an example the intersection at
$\sin\theta_2=\sin\theta_3=0$, the other two pairs having a similar
energy-momentum tensor.

Finally, and thanks to the third Lovelock term, there is a non-trivial
contribution to the generalised Einstein tensor that allows us to
match two codimension 6 branes at the 
intersection of the three codimension 2 branes (\textit{i.e.} at the
points $\theta_1=\theta_2=0$ and $\theta_3=0$ or $\theta_3=\pi$) with
the following energy-momentum tensor 
\begin{equation}
T^{\mathrm{Cod}\,6}_{\mu\nu}=
24 \alpha_3 
\eta_{\mu\nu}(1-\beta_1)(1-\beta_2)(1-\beta_3)
\frac{\delta(\sin\theta_1)}{\sqrt{k_1}}
\frac{\delta(\sin\theta_2)}{\sqrt{k_2}}
\frac{\delta(\sin\theta_3)}{\sqrt{k_3}}.
\end{equation}
Again our general discussion tells us that the fourth Lovelock term contributes to the 3-brane matching condition a term proportional to the Einstein tensor for the induced metric on the brane. In
our case this tensor does of course vanish as corresponds to flat
space. Notice that the energy-momentum tensors of all the branes are fixed in terms of the deficit angles and $R$, since we are imposing a particular background. So one should fine tune the brane tensions and the $vevs$ of the brane fluxes in order to find this solution.

\section{Conclusions}

In Einstein gravity there are no regular solutions representing
isolated sources of codimension higher than 2. This means that
classical solutions of higher dimensional gravity representing branes
of codimension three or higher are singular when we model the brane
as a delta-like contribution to the energy-momentum tensor. So in
order to extract any useful information about the behaviour of gravity
on or close to the brane we have to go beyond this approximation
($i.e.$ to a theory in which the brane has finite width and some
internal structure and/or to a theory that resolves the singularities
appearing in classical gravity). In this letter, generalising the
ideas presented in \cite{Lee:2004hh}, we have shown that when we
consider all available Lanczos-Lovelock terms in higher dimensions, it
is possible to build solutions representing infinitesimally thin
branes of higher codimension living in the intersection branes of
lower codimension, in a background that is otherwise free of
singularities, and in particular where the induced metric in all the
branes is well defined. The building blocks of this constructions are
codimension 1 and 2 branes, since it is known how to generate isolated
one- and two-dimensional delta functions in the Riemann tensor, while
the higher dimensional deltas are obtained in the higher order
Lovelock terms as the product of the ones appearing in the Riemann
tensor. The structure of the Lanczos-Lovelock Lagrangians, and in
particular the quasilinearity of the equations with respect to the
second derivatives of the metric, ensures that singularities of the
type $\delta(y)^2$ are not present in the generalised Einstein
tensor. Moreover, the delta functions only appear in the components of
this tensor along the brane dimensions, and thus this solutions have a
natural interpretation as braneworlds. 

We have analysed the structure of the matching conditions for the
intersection of codimension 1 and/or codimension 2 branes. For the
codimension 2 case, under the assumption that the curvature in the
bulk does not diverge as we approach the brane, this equations are
remarkably simple, since they imply that the induced metric and matter
on the brane satisfy the Lovelock equations corresponding with its
dimensionality. In particular one obtains 4D Einstein gravity in the
case of a 3-brane embedded in a $2n-$dimensional spacetime. We have
presented two explicit examples, the intersection of a 5-brane with a
4-brane in 7D, yielding a codimension 3 brane, and the intersection of
three 7-branes in 10D, yielding three 5-branes and two 3-branes in its
intersections. All of these branes have a non-trivial energy-momentum
tensor and its effect in the background is fully taken into account. 

Our results open up the possibility of studying cosmology and other
gravity related phenomena in ten dimensional intersecting braneworlds within the framework of classical gravity, and provide the first example of
non-singular higher codimension braneworlds where the backreaction of
the branes in the background is fully taken into account.

\section*{Acknowledgements}
It is a pleasure to thank S. Abel and R. Gregory for useful
discussions. This work has been funded by PPARC.


\end{document}